\documentclass[preprint, 3p,
authoryear]{elsarticle} %review=doublespace preprint=single 5p=2 column
\usepackage[hyphens]{url}

  \journal{Arxiv} % Sets Journal name

\usepackage{graphicx}
\usepackage[T1]{fontenc}
\usepackage{lmodern}
\usepackage{amssymb,amsmath}
\usepackage{lineno} % add
\usepackage{ifxetex,ifluatex}
\usepackage{fixltx2e} % provides \textsubscript
\IfFileExists{upquote.sty}{\usepackage{upquote}}{}
\ifnum 0\ifxetex 1\fi\ifluatex 1\fi=0 % if pdftex
  \usepackage[utf8]{inputenc}
\else % if luatex or xelatex
  \usepackage{fontspec}
  \ifxetex
    \usepackage{xltxtra,xunicode}
  \fi
  \defaultfontfeatures{Mapping=tex-text,Scale=MatchLowercase}
  
\fi
\IfFileExists{microtype.sty}{\usepackage{microtype}}{}
\usepackage[]{natbib}
\ifxetex
  \usepackage[setpagesize=false, % page size defined by xetex
              unicode=false, % unicode breaks when used with xetex
              xetex]{hyperref}
\else
  \usepackage[unicode=true]{hyperref}
\fi
\hypersetup{breaklinks=true,
            bookmarks=true,
            pdfauthor={},
            pdftitle={Alzheimer's Clinical Research Data via R Packages: the alzverse},
            colorlinks=false,
            urlcolor=blue,
            linkcolor=magenta,
            pdfborder={0 0 0}}

\usepackage{color}
\usepackage{fancyvrb}

\DefineVerbatimEnvironment{Highlighting}{Verbatim}{commandchars=\\\{\}}
\usepackage{framed}
\definecolor{shadecolor}{RGB}{248,248,248}
\newenvironment{Shaded}{\begin{snugshade}}{\end{snugshade}}

\newcommand{\AttributeTok}[1]{\textcolor[rgb]{0.13,0.29,0.53}{#1}}

\newcommand{\CommentTok}[1]{\textcolor[rgb]{0.56,0.35,0.01}{\textit{#1}}}

\newcommand{\ConstantTok}[1]{\textcolor[rgb]{0.56,0.35,0.01}{#1}}

\newcommand{\DecValTok}[1]{\textcolor[rgb]{0.00,0.00,0.81}{#1}}

\newcommand{\FloatTok}[1]{\textcolor[rgb]{0.00,0.00,0.81}{#1}}
\newcommand{\FunctionTok}[1]{\textcolor[rgb]{0.13,0.29,0.53}{\textbf{#1}}}

\newcommand{\NormalTok}[1]{#1}

\newcommand{\OtherTok}[1]{\textcolor[rgb]{0.56,0.35,0.01}{#1}}

\newcommand{\SpecialCharTok}[1]{\textcolor[rgb]{0.81,0.36,0.00}{\textbf{#1}}}

\newcommand{\StringTok}[1]{\textcolor[rgb]{0.31,0.60,0.02}{#1}}

\providecommand{\tightlist}{%
  \setlength{\itemsep}{0pt}\setlength{\parskip}{0pt}}

\begin{document}

\begin{frontmatter}

  \title{Alzheimer's Clinical Research Data via R Packages: the
alzverse}
    \author[ATRI]{Michael C. Donohue%
  \corref{cor1}%
  }
  
    \author[ATRI]{Kedir Hussen%
  }
  
    \author[ATRI]{Oliver Langford%
  }
  
    \author[ATRI]{Richard Gallardo%
  }
  
    \author[ATRI]{Gustavo Jimenez-Maggiora%
  }
  
    \author[ATRI]{Paul S. Aisen%
  }
  
    \author[]{for the Alzheimer's Disease Neuroimaging Initiative%
  \fnref{1}}
  
      \affiliation[ATRI]{
    organization={Alzheimer's Therapeutic Research Institute, University
of Southern California},city={San
Diego},postcode={92121},state={CA},country={United States},}
    \cortext[cor1]{Corresponding author}
    \fntext[1]{Data used in preparation of this article were obtained
from the Alzheimer's Disease Neuroimaging Initiative (ADNI) database
(\url{adni.loni.usc.edu}). As such, the investigators within the ADNI
contributed to the design and implementation of ADNI and/or provided
data but did not participate in the analysis or writing of this report.
A complete listing of ADNI investigators can be found at:
\url{http://adni.loni.usc.edu/wp-content/uploads/how_to_apply/ADNI_Acknowledgement_List.pdf}.}
  
  \begin{abstract}
  Sharing clinical research data is essential for advancing research in
  Alzheimer's disease (AD) and other therapeutic areas. However,
  challenges in data accessibility, standardization, documentation,
  usability, and reproducibility continue to impede this goal. In this
  article, we highlight the advantages of using R packages to overcome
  these challenges using two examples. The \texttt{A4LEARN} R package
  includes data from a randomized trial (the Anti-Amyloid Treatment in
  Asymptomatic Alzheimer's {[}A4{]} study) and its companion
  observational study of biomarker negative individuals (the
  Longitudinal Evaluation of Amyloid Risk and Neurodegeneration
  {[}LEARN{]} study). The \texttt{ADNIMERGE2} R package includes data
  from the Alzheimer's Disease Neuroimaging Initiative (ADNI), a
  longitudinal observational biomarker and imaging study. These packages
  collect data, documentation, and reproducible analysis vignettes into
  a portable bundle that can be installed and browsed within commonly
  used R programming environments. We also introduce the
  \texttt{alzverse} package which leverages a common data standard to
  combine study-specific data packages to facilitate meta-analyses. By
  promoting collaboration, transparency, and reproducibility, R data
  packages can play a vital role in accelerating clinical research.
  \end{abstract}
  
 \end{frontmatter}

\subsection{Introduction}\label{introduction}

Alzheimer's disease (AD) is among the most important neurodegenerative
diseases worldwide. The Alzheimer's Disease Neuroimaging Initiative
(ADNI) \citep{weiner2025overview} and similar projects have accumulated
vast quantities of clinical, neuroimaging, and biomarker data, creating
opportunities for scientific advances in the understanding and treatment
of AD. ADNI data have supported more than 6000 scientific papers
publications \citep{weiner2025overview}.

Availability of such data is on the rise, due in part to data sharing
mandates from funders like the National Institutes of Health. However,
it often takes considerable time and effort for researchers to gain
sufficient familiarity with the data to produce meaningful analyses.
Learning curves can be steep due to inadequate or hard-to-locate
documentation and example analysis code.

Open-source software solutions, particularly in the form of R packages
\citep{R-base, wickham2023r}, offer significant potential to address
these barriers. The R programming language is widely used in
biostatistics, machine learning, and clinical research. R packages are a
well-known means for distributing cutting edge statistical software and
documentation, and they often include data and analysis vignettes which
demonstrate how the methods can be applied. Importantly, R packages can
also be an effective tool to share data itself. For example, the authors
have maintained the \texttt{ADNIMERGE} R data package since 2017
\citep{ADNI2023}. It has been cited by about 250 articles\footnote{Based
  on a Google Scholar search of ``ADNIMERGE'' AND ``package''} and has
inspired related projects, such as the ANMERGE package of AddNeuroMed
Consortium data \citep{birkenbihl2021anmerge}. Similarly,
\citet{vuorre2021sharing} demonstrated the utility of R packages for
sharing data and analysis code from psychological experiments.

We discuss how R packages can also facilitate easy access,
harmonization, and analysis of larger clinical datasets from AD studies.
These packages are built with the goal of providing an audit trail of
derived data provenance, supporting reproducible research, and
leveraging outstanding R tools for unit testing and validation
\citep{testthat}, websites \citep{pkgdown}, and regulatory submissions
\citep{pharmaverse2023}.

In this paper we discuss the advantages of using R packages to share
large clinical study datasets, and provide two new examples: the
\texttt{A4LEARN} packages which includes data from a randomized trial
(the Anti-Amyloid Treatment in Asymptomatic Alzheimer's {[}A4{]} study)
\citep{sperling2023trial} and its companion observational study of
biomarker negative individuals (the Longitudinal Evaluation of Amyloid
Risk and Neurodegeneration {[}LEARN{]} study)
\citep{sperling2024amyloid}; and the \texttt{ADNIMERGE2} package, which
includes latest data from the Alzheimer's Disease Neuroimaging
Initiative (ADNI) \citep{weiner2025overview}. We also introduce the
\href{https://atri-biostats.github.io/alzverse/}{\texttt{alzverse}} meta
data package, which demonstrates how \texttt{ADNIMERGE2},
\texttt{A4LEARN}, and other R data packages can be combined to
facilitate meta-analyses.

\subsection{Advantages of R Data
Packages}\label{advantages-of-r-data-packages}

\subsubsection{Reproducibility, Portability and
Documentation}\label{reproducibility-portability-and-documentation}

The most important advantage of using R data packages for sharing
clinical research data is that it facilitates reproducible research. R
is widely available and free to download \citep{R}, commonly used in
statistics courses, and has active development communities including
academic statisticians, pharmaceutical statisticians, and commercial
enterprises such as Posit's RStudio Integrated Development Environment
(IDE).

All R packages include a manual that details the functions and datasets
in a standardized format. This content is linked to the R object's name,
is readily accessible to the R user (e.g.~by typing \texttt{?t.test}, or
\texttt{?cars}) and can be browsed within the IDE. This is a substantial
improvement over the typical case in which relevant data documentation
might be scattered across unlinked documents or data dictionary
spreadsheets. R packages also typically provide analysis ``vignettes'',
which demonstrate how functions can be applied to available datasets to
produce analysis results as tables or figures. These are also linked and
can be browsed within the IDE. Our \texttt{A4LEARN} package contains a
vignette to exactly reproduce key findings of the published manuscript
reporting the trial results \citep{sperling2023trial}. This code can be
used by outside researchers to jump start their own inquiries and help
ensure the data is being used correctly and efficiently in a manner
consistent with the intentions of the study team.

The R package bundle of data, R functions, documentation and vignettes
is made portable as an efficiently compressed file which can be
installed on any machine running R. Data files within the package are
also efficiently compressed using R's \texttt{.RData} file format. These
\texttt{.RData} files can be read by SPSS, Stata, and SAS, allowing
researchers to access and analyze the data using a variety of
statistical software beyond R. Another advantage of \texttt{.RData}
compared to tabular text files (e.g.~\texttt{.csv} files), is that they
can utilize R object classes such as dates and factor variables,
eliminating the need to process and annotate data prior to analysis.

The \texttt{pkgdown} R package \citep{pkgdown} makes it trivial to
export documentation and vignettes as a website, and integrates well
with code repositories such as GitHub. These \texttt{pkgdown} websites
are searchable and provide access to documentation and examples for
researchers, even if they do not use R. See
\href{https://atri-biostats.github.io/A4LEARN}{atri-biostats.github.io/A4LEARN/}
and
\href{https://atri-biostats.github.io/ADNIMERGE2}{atri-biostats.github.io/ADNIMERGE2}
for examples.

\subsubsection{Standardized and Efficient Workflow and
Testing}\label{standardized-and-efficient-workflow-and-testing}

Questions often arise about the original source of data or how derived
variables were determined. Therefore, it is crucial to preserve a record
of data provenance. This is important whenever reproducible research is
a goal, and particularly for submissions to regulators, such as the FDA.
The standardized R package structure and build workflow makes the data
provenance transparent. R packages also have a standardized framework
for testing \citep{testthat} and tools for ``assertive'' programming to
verify assumptions about the data \citep{assertr}.

The R package structure and workflow have been well-documented
\citep{wickham2023r}. Here we briefly review the structure while
highlighting some key aspects in the context of the clinical research
data.

\textbf{data-raw}. The \texttt{data-raw} directory is intended to
contain raw data and code to import and process raw data and store
\texttt{.Rdata} files in the \texttt{data} directory. Raw data can be
preserved in the package with minimal manipulation, or it can be
processed attaching meta data (variable labels and units), and ensuring
factors and dates are stored as the correct object class. Data
dictionary spreadsheets can be parsed to provide content for manual
pages \citep{roxygen2}.

\textbf{vignettes}. The \texttt{vignettes} directory contains analysis
demonstrations, typically as Rmarkdown (\texttt{.Rmd}) files
\citep{rmarkdown}. We prefer to create derived datasets and variables in
the form of a vignette as well, so that derivations are easily
accessible to researchers within the IDE. These vignettes can include
assertive programming to ensure data conforms to expectations
\citep{assertr}. \texttt{ADNIMERGE2} contains vignettes which use
\href{https://pharmaverse.org/}{\texttt{pharmaverse}} workflows to
derive Clinical Data Interchange Standards Consortium (CDISC) Analysis
Data Model (ADaM) datasets.

\textbf{R}. The R directory contains \texttt{.R} files with code
defining R functions and their manual content \citep{roxygen2}. This
directory can store scoring functions, which might be necessary to
derive scores from item-level data from psychometric assessments, for
example.

\textbf{testthat}. The \texttt{testthat} directory includes automated
tests that are checked when the package is built. Critically important
code that merits replication by independent programmers can be tested
here to ensure equivalent results with real or test data.

\textbf{reports}. Report code that is not desired as a vignette can be
stored in a separate directory for general reports. Of note,
\texttt{rmarkdown} supports ``parameterized reports'', which can produce
different output depending on the supplied parameter(s). Clinical trials
like A4 often include several outcomes collected on the same schedule
and analyzed with the same approach. Instead of writing identical code
for several outcomes, one generic parameterized rmarkdown file can
produce all these reports. Clinical trial outcomes are often aggregated
into one long dataset with a row for each subject, time point, and
outcome (see \texttt{ADQS} in the \texttt{A4LEARN} package). The
parameterized report can filter this long dataset for the desired
outcome and analyze only that outcome. Furthermore, using R parallel
programming tools (e.g. \citet{multidplyr}), these reports can be
produced in parallel. In the case of the A4 trial read out, once data
from the blinded phase was locked and unblinded it only took about 30
minutes to build the final data package and render all planned analysis
reports and summary slide decks. Prior to data lock, code and output
were tested and reviewed using pseudo arm codes.

\subsection{Example R Data Packages}\label{example-r-data-packages}

The primary data sources for the packages are derived from the
Alzheimer's Disease Neuroimaging Initiative (ADNI) and the A4 and LEARN
companion studies \citep{jimenez2024maximizing}. These datasets include
longitudinal clinical and neuroimaging data, cognitive test scores,
genetic and biomarker data, and other modalities that have been
harmonized to facilitate cross-study comparisons.

To ensure usability and consistency, we curated the datasets by mapping
variables to standardized terminologies (e.g., CDISC ADaM), handling
missing data through imputation techniques, and deriving key analysis
variables.

\texttt{A4LEARN} and \texttt{ADNIMERGE2} were developed using best
practices in R package development, including:

\begin{itemize}
\tightlist
\item
  R Package Architecture: Each package is modular, supporting various
  stages of data analysis, from raw data processing to the generation of
  regulatory-compliant datasets.
\item
  Data Standardization: The packages support standardization of clinical
  data, with built-in functions to harmonize variable formats, handle
  missing values, and generate standardized metadata.
\item
  Reproducibility: Built-in vignettes and examples guide users through
  the installation, data loading, and analysis processes. The packages
  integrate with tools like \texttt{renv} and Docker to facilitate
  reproducibility in different computing environments.
\end{itemize}

The R package framework facilitates the creation of ADaM datasets, which
are the gold standard for statistical analysis in clinical trials. The
\texttt{admiral} package \citep{admiral} is used to generate ADaM
datasets for \texttt{ADNIMERGE2}. Other \texttt{pharmaverse} tools can
be used to create regulatory-compliant tables, listings, and figures. R
code to produce the example summaries below is included in the Code
Availability section.

The \texttt{ADNIMERGE2} package was built using \texttt{pharmaverse}
tools. It can be downloaded from \url{loni.usc.edu}. Below are examples
of some basic summaries of participant characteristics by phase, or
wave, of ADNI can be created using the \texttt{ADSL} table (Table 1) and
a spaghetti plot of ADAS-Cog scores using the \texttt{ADQS} table
(Figure 1). Similarly, \texttt{A4LEARN} package data can be easily
summarized using its \texttt{SUBJINFO} table (Table 2). Finally, data
from all three studies can be easily summarized using the meta
\texttt{ADQS} table of the \texttt{alzverse} package (Figure 2).

\includegraphics[width=5in,height=9in]{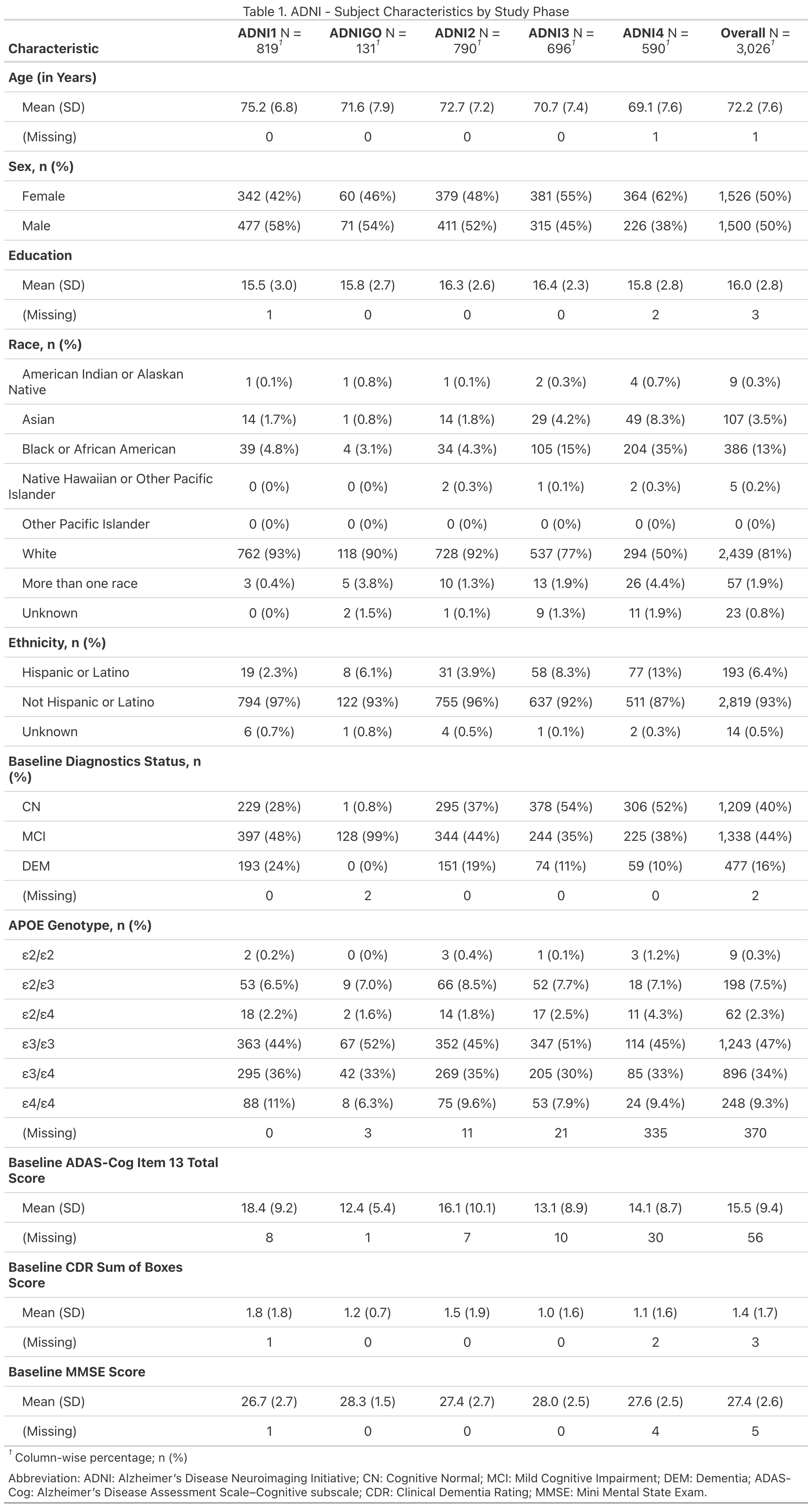}

\begin{figure}
\includegraphics[width=1\linewidth,alt={Individual profile plot}]{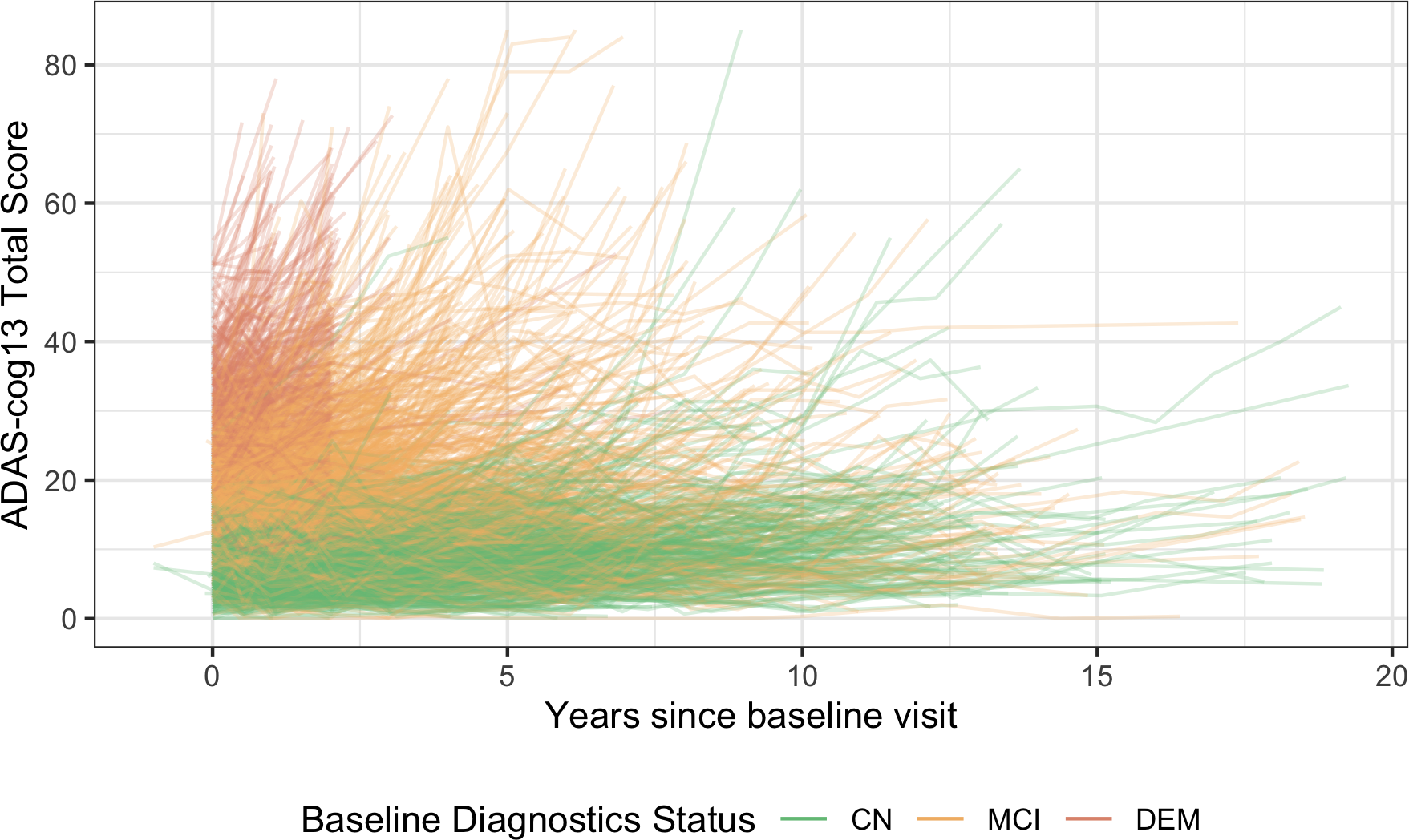} \caption{Figure 1. Spaghetti plot of ADAS-cog13 scores in ADNI by baseline clinical diagnosis.}\label{fig:adni-adas-indiv-profile-plot}
\end{figure}

\includegraphics[width=4.5in,height=5.22in]{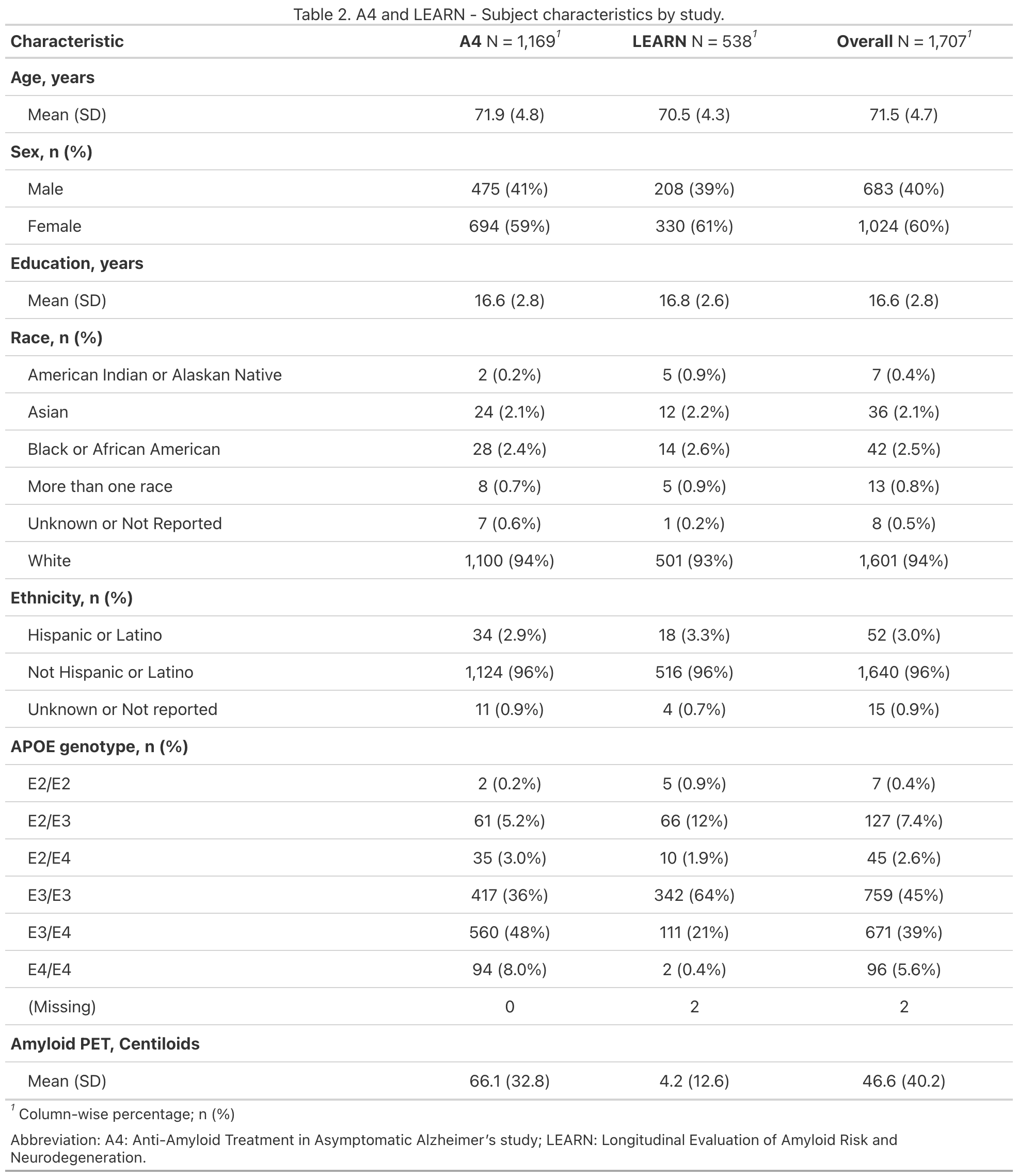}

\begin{figure}
\includegraphics[width=1\linewidth,alt={Individual profile plot}]{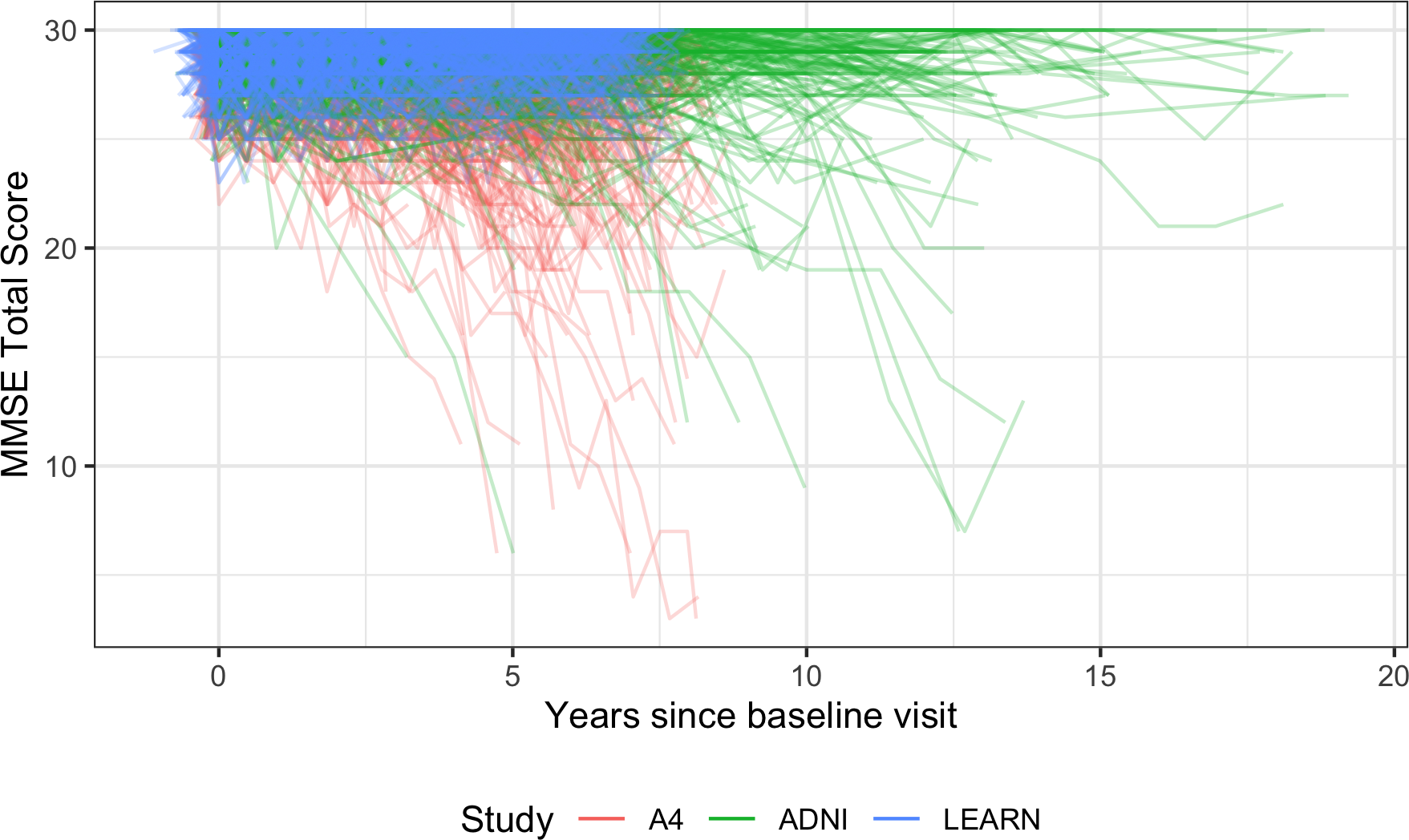} \caption{Figure 2. Spaghetti plot of MMSE scores in ADNI CN, A4, and LEARN.}\label{fig:adni-a4-learn-indiv-profile-plot}
\end{figure}

\subsection{Discussion}\label{discussion}

The development of \texttt{A4LEARN} and \texttt{ADNIMERGE2} represents a
step forward in enabling the Alzheimer's disease research community to
share and analyze data more effectively, serving as a template for
additional future study packages. These packages facilitate the
transition from proprietary software like SAS to open-source tools,
allowing greater flexibility and transparency in the research process.
The shift from SAS to R reflects a broader trend in the clinical
research community toward open-source and reproducible research
practices, as exemplified by the \texttt{pharmaverse} project.

Challenges remain, particularly in the area of data access. In our
examples data packages can be sourced using the existing data access
models. This puts the onus on data users to go to different sites to
obtain data. Once retrieved, the use of locally installed packages poses
potential risks, which can be mitigated by using containerization and
package management tools like Docker and \texttt{renv} \citep{renv} for
version control. A streamlined future improvement on this model might
entail the use access keys and Application Programming Interfaces (APIs)
to source data directly within R.

We envision the \texttt{alzverse} project expanding to include many more
studies. Our plan is to continue building R data packages for studies
conducted by the Alzheimer's Therapeutic Research Institute and the
Alzheimer's Clinical Trial Consortium, with the hope of eventually
expanding to other groups through a larger collaborative community. At
some point, however, the meta-package might become impractically large.
One way to mitigate this limitation would be to allow users to select
which studies to include in their personalized version of the
\texttt{alzverse}.

\subsection{Data Availability}\label{data-availability}

Data is available from:

\begin{itemize}
\tightlist
\item
  \texttt{A4LEARN}: \url{A4StudyData.org}
\item
  \texttt{ADNIMERGE2}: \url{loni.usc.edu}
\end{itemize}

Documentation is available from

\begin{itemize}
\tightlist
\item
  \texttt{A4LEARN}:
  \href{https://atri-biostats.github.io/A4LEARN}{atri-biostats.github.io/A4LEARN/}
\item
  \texttt{ADNIMERGE2}:
  \href{https://atri-biostats.github.io/ADNIMERGE2}{atri-biostats.github.io/ADNIMERGE2}
\end{itemize}

\subsection{Code Availability}\label{code-availability}

R code is available for download from the following repositories:

\begin{itemize}
\tightlist
\item
  alzverse:
  \href{https://github.com/atri-biostats/alzverse}{github.com/atri-biostats/alzverse}
\item
  A4LEARN:
  \href{https://github.com/atri-biostats/A4LEARN}{github.com/atri-biostats/A4LEARN}
\item
  ADNIMERGE2:
  \href{https://github.com/atri-biostats/ADNIMERGE2}{github.com/atri-biostats/ADNIMERGE2}
\end{itemize}

\subsubsection{R Code for the ADNIMERGE2
Example}\label{r-code-for-the-adnimerge2-example}

\begin{Shaded}
\begin{Highlighting}[]
\NormalTok{tab1 }\OtherTok{\textless{}{-}} \FunctionTok{tbl\_summary}\NormalTok{(}
  \AttributeTok{data =}\NormalTok{ ADNIMERGE2}\SpecialCharTok{::}\NormalTok{ADSL }\SpecialCharTok{\%\textgreater{}\%} \FunctionTok{filter}\NormalTok{(ENRLFL }\SpecialCharTok{\%in\%} \StringTok{"Y"}\NormalTok{),}
  \AttributeTok{by =}\NormalTok{ ORIGPROT,}
  \AttributeTok{include =} \FunctionTok{c}\NormalTok{(AGE, SEX, EDUC, RACE, ETHNIC, DX, APOE, ADASTT13, CDRSB, MMSCORE),}
  \AttributeTok{type =} \FunctionTok{all\_continuous}\NormalTok{() }\SpecialCharTok{\textasciitilde{}} \StringTok{"continuous2"}\NormalTok{,}
  \AttributeTok{statistic =} \FunctionTok{list}\NormalTok{(}
    \FunctionTok{all\_continuous}\NormalTok{() }\SpecialCharTok{\textasciitilde{}} \StringTok{"\{mean\} (\{sd\})"}\NormalTok{,}
    \FunctionTok{all\_categorical}\NormalTok{() }\SpecialCharTok{\textasciitilde{}} \StringTok{"\{n\} (\{p\}\%)"}
\NormalTok{  ),}
  \AttributeTok{digits =} \FunctionTok{all\_continuous}\NormalTok{() }\SpecialCharTok{\textasciitilde{}} \DecValTok{1}\NormalTok{,}
  \AttributeTok{percent =} \StringTok{"column"}\NormalTok{,}
  \AttributeTok{missing\_text =} \StringTok{"(Missing)"}
\NormalTok{) }\SpecialCharTok{\%\textgreater{}\%}
  \FunctionTok{add\_overall}\NormalTok{(}\AttributeTok{last =} \ConstantTok{TRUE}\NormalTok{) }\SpecialCharTok{\%\textgreater{}\%}
  \FunctionTok{add\_stat\_label}\NormalTok{(}\AttributeTok{label =} \FunctionTok{all\_continuous2}\NormalTok{() }\SpecialCharTok{\textasciitilde{}} \StringTok{"Mean (SD)"}\NormalTok{) }\SpecialCharTok{\%\textgreater{}\%}
  \FunctionTok{modify\_footnote\_header}\NormalTok{(}
    \AttributeTok{footnote =} \StringTok{"Column{-}wise percentage; n (\%)"}\NormalTok{,}
    \AttributeTok{columns =} \FunctionTok{all\_stat\_cols}\NormalTok{(),}
    \AttributeTok{replace =} \ConstantTok{TRUE}
\NormalTok{  ) }\SpecialCharTok{\%\textgreater{}\%}
  \FunctionTok{modify\_abbreviation}\NormalTok{(}\AttributeTok{abbreviation =} 
  \StringTok{"ADNI: Alzheimer’s Disease Neuroimaging Initiative; }
\StringTok{    CN: Cognitive Normal; MCI: Mild Cognitive Impairment; }
\StringTok{    DEM: Dementia; }
\StringTok{    ADAS{-}Cog: Alzheimer\textquotesingle{}s Disease Assessment Scale{-}{-}Cognitive subscale; }
\StringTok{    CDR: Clinical Dementia Rating; }
\StringTok{    MMSE: Mini Mental State Exam."}\NormalTok{) }\SpecialCharTok{\%\textgreater{}\%}
  \FunctionTok{modify\_caption}\NormalTok{(}
    \AttributeTok{caption =} \StringTok{"Table 1. ADNI {-} Subject Characteristics by Study Phase"}\NormalTok{) }\SpecialCharTok{\%\textgreater{}\%}
  \FunctionTok{bold\_labels}\NormalTok{() }\SpecialCharTok{\%\textgreater{}\%}
  \FunctionTok{as\_gt}\NormalTok{()}

\FunctionTok{gtsave}\NormalTok{(tab1, }\AttributeTok{filename =} \StringTok{"Table{-}1{-}ADNI.png"}\NormalTok{)}
\NormalTok{knitr}\SpecialCharTok{::}\FunctionTok{include\_graphics}\NormalTok{(}\StringTok{"Table{-}1{-}ADNI.png"}\NormalTok{)}
\end{Highlighting}
\end{Shaded}

\begin{Shaded}
\begin{Highlighting}[]
\CommentTok{\# Individual profile (spaghetti) plot}
\NormalTok{ADNIMERGE2}\SpecialCharTok{::}\NormalTok{ADQS }\SpecialCharTok{\%\textgreater{}\%}
  \CommentTok{\# Enrolled participant}
  \FunctionTok{filter}\NormalTok{(ENRLFL }\SpecialCharTok{\%in\%} \StringTok{"Y"}\NormalTok{) }\SpecialCharTok{\%\textgreater{}\%}
  \CommentTok{\# ADAS{-}cog item{-}13 total score}
  \FunctionTok{filter}\NormalTok{(PARAMCD }\SpecialCharTok{\%in\%} \StringTok{"ADASTT13"}\NormalTok{) }\SpecialCharTok{\%\textgreater{}\%}
  \FunctionTok{mutate}\NormalTok{(}\AttributeTok{Years =} \FunctionTok{convert\_number\_days}\NormalTok{(ADY, }\AttributeTok{unit =} \StringTok{\textquotesingle{}year\textquotesingle{}}\NormalTok{)) }\SpecialCharTok{\%\textgreater{}\%}
  \FunctionTok{filter}\NormalTok{(}\SpecialCharTok{!}\FunctionTok{if\_any}\NormalTok{(}\FunctionTok{all\_of}\NormalTok{(}\FunctionTok{c}\NormalTok{(}\StringTok{"Years"}\NormalTok{, }\StringTok{"DX"}\NormalTok{, }\StringTok{"AVAL"}\NormalTok{)), }\SpecialCharTok{\textasciitilde{}} \FunctionTok{is.na}\NormalTok{(.x))) }\SpecialCharTok{\%\textgreater{}\%}
\FunctionTok{ggplot}\NormalTok{(}\FunctionTok{aes}\NormalTok{(}\AttributeTok{x =}\NormalTok{ Years, }\AttributeTok{y =}\NormalTok{ AVAL, }\AttributeTok{group =}\NormalTok{ USUBJID, }\AttributeTok{color =}\NormalTok{ DX)) }\SpecialCharTok{+}
  \FunctionTok{geom\_line}\NormalTok{(}\AttributeTok{alpha =} \FloatTok{0.25}\NormalTok{) }\SpecialCharTok{+}
  \FunctionTok{scale\_color\_manual}\NormalTok{(}\AttributeTok{values =} \FunctionTok{c}\NormalTok{(}\StringTok{"\#73C186"}\NormalTok{, }\StringTok{"\#F2B974"}\NormalTok{, }\StringTok{"\#DF957C"}\NormalTok{, }\StringTok{"\#999999"}\NormalTok{)) }\SpecialCharTok{+}
  \FunctionTok{labs}\NormalTok{(}
    \AttributeTok{y =} \StringTok{"ADAS{-}cog13 Total Score"}\NormalTok{,}
    \AttributeTok{x =} \StringTok{"Years since baseline visit"}\NormalTok{,}
    \AttributeTok{color =} \StringTok{"Baseline Diagnostics Status"}\NormalTok{) }\SpecialCharTok{+}
  \FunctionTok{theme}\NormalTok{(}\AttributeTok{legend.position =} \StringTok{"bottom"}\NormalTok{) }\SpecialCharTok{+}
  \FunctionTok{guides}\NormalTok{(}\AttributeTok{colour =} \FunctionTok{guide\_legend}\NormalTok{(}\AttributeTok{override.aes =} \FunctionTok{list}\NormalTok{(}\AttributeTok{alpha =} \DecValTok{1}\NormalTok{)))}
\end{Highlighting}
\end{Shaded}

\subsubsection{R Code for the A4LEARN
Example}\label{r-code-for-the-a4learn-example}

\begin{Shaded}
\begin{Highlighting}[]
\NormalTok{tab2 }\OtherTok{\textless{}{-}} \FunctionTok{tbl\_summary}\NormalTok{(}
  \AttributeTok{data =}\NormalTok{ A4LEARN}\SpecialCharTok{::}\NormalTok{SUBJINFO }\SpecialCharTok{\%\textgreater{}\%} \FunctionTok{filter}\NormalTok{(SUBSTUDY }\SpecialCharTok{\%in\%} \FunctionTok{c}\NormalTok{(}\StringTok{\textquotesingle{}A4\textquotesingle{}}\NormalTok{, }\StringTok{\textquotesingle{}LEARN\textquotesingle{}}\NormalTok{)),}
  \AttributeTok{by =}\NormalTok{ SUBSTUDY,}
  \AttributeTok{include =} \FunctionTok{c}\NormalTok{(AGEYR, SEX, EDCCNTU, RACE, ETHNIC, APOEGN, AMYLCENT),}
  \AttributeTok{label =} \FunctionTok{list}\NormalTok{(}\AttributeTok{AGEYR =} \StringTok{"Age, years"}\NormalTok{, }\AttributeTok{SEX =} \StringTok{\textquotesingle{}Sex\textquotesingle{}}\NormalTok{, }\AttributeTok{EDCCNTU =} \StringTok{\textquotesingle{}Education, years\textquotesingle{}}\NormalTok{, }
    \AttributeTok{RACE =} \StringTok{\textquotesingle{}Race\textquotesingle{}}\NormalTok{, }\AttributeTok{ETHNIC =} \StringTok{\textquotesingle{}Ethnicity\textquotesingle{}}\NormalTok{, }\AttributeTok{APOEGN =} \StringTok{\textquotesingle{}APOE genotype\textquotesingle{}}\NormalTok{,}
    \AttributeTok{AMYLCENT =} \StringTok{\textquotesingle{}Amyloid PET, Centiloids\textquotesingle{}}\NormalTok{),}
  \AttributeTok{type =} \FunctionTok{all\_continuous}\NormalTok{() }\SpecialCharTok{\textasciitilde{}} \StringTok{"continuous2"}\NormalTok{,}
  \AttributeTok{statistic =} \FunctionTok{list}\NormalTok{(}
    \FunctionTok{all\_continuous}\NormalTok{() }\SpecialCharTok{\textasciitilde{}} \StringTok{"\{mean\} (\{sd\})"}\NormalTok{,}
    \FunctionTok{all\_categorical}\NormalTok{() }\SpecialCharTok{\textasciitilde{}} \StringTok{"\{n\} (\{p\}\%)"}
\NormalTok{  ),}
  \AttributeTok{digits =} \FunctionTok{all\_continuous}\NormalTok{() }\SpecialCharTok{\textasciitilde{}} \DecValTok{1}\NormalTok{,}
  \AttributeTok{percent =} \StringTok{"column"}\NormalTok{,}
  \AttributeTok{missing\_text =} \StringTok{"(Missing)"}
\NormalTok{) }\SpecialCharTok{\%\textgreater{}\%}
  \FunctionTok{add\_overall}\NormalTok{(}\AttributeTok{last =} \ConstantTok{TRUE}\NormalTok{) }\SpecialCharTok{\%\textgreater{}\%}
  \FunctionTok{add\_stat\_label}\NormalTok{(}\AttributeTok{label =} \FunctionTok{all\_continuous2}\NormalTok{() }\SpecialCharTok{\textasciitilde{}} \StringTok{"Mean (SD)"}\NormalTok{) }\SpecialCharTok{\%\textgreater{}\%}
  \FunctionTok{modify\_footnote\_header}\NormalTok{(}
    \AttributeTok{footnote =} \StringTok{"Column{-}wise percentage; n (\%)"}\NormalTok{,}
    \AttributeTok{columns =} \FunctionTok{all\_stat\_cols}\NormalTok{(),}
    \AttributeTok{replace =} \ConstantTok{TRUE}
\NormalTok{  ) }\SpecialCharTok{\%\textgreater{}\%}
  \FunctionTok{modify\_abbreviation}\NormalTok{(}\AttributeTok{abbreviation =} 
    \StringTok{"A4: Anti{-}Amyloid Treatment in Asymptomatic Alzheimer’s study; }
\StringTok{    LEARN: Longitudinal Evaluation of Amyloid Risk and Neurodegeneration."}\NormalTok{) }\SpecialCharTok{\%\textgreater{}\%}
  \FunctionTok{modify\_caption}\NormalTok{(}
    \AttributeTok{caption =} \StringTok{"Table 2. A4 and LEARN {-} Subject characteristics by study."}\NormalTok{) }\SpecialCharTok{\%\textgreater{}\%}
  \FunctionTok{bold\_labels}\NormalTok{() }\SpecialCharTok{\%\textgreater{}\%}
  \FunctionTok{as\_gt}\NormalTok{()}

\FunctionTok{gtsave}\NormalTok{(tab2, }\AttributeTok{filename =} \StringTok{"Table{-}2{-}A4LEARN.png"}\NormalTok{)}
\NormalTok{knitr}\SpecialCharTok{::}\FunctionTok{include\_graphics}\NormalTok{(}\StringTok{"Table{-}2{-}A4LEARN.png"}\NormalTok{)}
\end{Highlighting}
\end{Shaded}

\subsubsection{R Code for the alzverse
Example}\label{r-code-for-the-alzverse-example}

\begin{Shaded}
\begin{Highlighting}[]
\NormalTok{alzverse}\SpecialCharTok{::}\NormalTok{ADQS }\SpecialCharTok{\%\textgreater{}\%}
  \FunctionTok{filter}\NormalTok{(DX }\SpecialCharTok{==} \StringTok{\textquotesingle{}CN\textquotesingle{}} \SpecialCharTok{|}\NormalTok{ STUDYID }\SpecialCharTok{\%in\%} \FunctionTok{c}\NormalTok{(}\StringTok{\textquotesingle{}A4\textquotesingle{}}\NormalTok{, }\StringTok{\textquotesingle{}LEARN\textquotesingle{}}\NormalTok{), }
\NormalTok{    PARAMCD }\SpecialCharTok{==} \StringTok{\textquotesingle{}MMSE\textquotesingle{}}\NormalTok{) }\SpecialCharTok{\%\textgreater{}\%}
  \FunctionTok{mutate}\NormalTok{(}\AttributeTok{Years =} \FunctionTok{convert\_number\_days}\NormalTok{(ADY, }\AttributeTok{unit =} \StringTok{\textquotesingle{}year\textquotesingle{}}\NormalTok{)) }\SpecialCharTok{\%\textgreater{}\%}
\FunctionTok{ggplot}\NormalTok{(}\FunctionTok{aes}\NormalTok{(}\AttributeTok{x =}\NormalTok{ Years, }\AttributeTok{y =}\NormalTok{ AVAL, }\AttributeTok{color =}\NormalTok{ STUDYID)) }\SpecialCharTok{+}
  \FunctionTok{geom\_line}\NormalTok{(}\FunctionTok{aes}\NormalTok{(}\AttributeTok{group =}\NormalTok{ USUBJID), }\AttributeTok{alpha =} \FloatTok{0.25}\NormalTok{) }\SpecialCharTok{+}
  \FunctionTok{labs}\NormalTok{(}
    \AttributeTok{y =} \StringTok{"MMSE Total Score"}\NormalTok{,}
    \AttributeTok{x =} \StringTok{"Years since baseline visit"}\NormalTok{,}
    \AttributeTok{color =} \StringTok{"Study"}\NormalTok{) }\SpecialCharTok{+}
  \FunctionTok{theme}\NormalTok{(}\AttributeTok{legend.position =} \StringTok{"bottom"}\NormalTok{) }\SpecialCharTok{+}
  \FunctionTok{guides}\NormalTok{(}\AttributeTok{colour =} \FunctionTok{guide\_legend}\NormalTok{(}\AttributeTok{override.aes =} \FunctionTok{list}\NormalTok{(}\AttributeTok{alpha =} \DecValTok{1}\NormalTok{)))}
\end{Highlighting}
\end{Shaded}

\subsection{Author Contributions}\label{author-contributions}

MCD contributed to conceptualization, software, validation, resources,
data curation, writing the original draft, writing review \& editing,
and supervision. KH, OL, and RG contributed to software, validation,
data curation, writing review \& editing. GJ-M contributed to
conceptualization, software, validation, resources, data curation,
writing review \& editing, and supervision. PSA contributed to
resources, writing review \& editing, and supervision.

\subsection{Competing Interests}\label{competing-interests}

MCD has research grants from NIH and the Alzheimer's Association, is a
consultant to F. Hoffmann-La Roche Ltd, and his spouse is a full-time
employee of Johnson \& Johnson.

PSA has research grants from NIH, the Alzheimer's Association, Janssen,
Lilly, and Eisai, and consults with Merck, Roche, Genen-tech, Abbvie,
Biogen, and ImmunoBrain Checkpoint.

Other authors report no relevant competing interests.

\subsection{Acknowledgements}\label{acknowledgements}

The authors are grateful to the ADNI, A4, and LEARN participants, their
families; and the study teams, including site investigators and staff.

Data collection and sharing for the Alzheimer's Disease Neuroimaging
Initiative (ADNI) is funded by the National Institute on Aging (NIA)
(National Institutes of Health {[}NIH{]} Grant U19AG024904). The grantee
organization is the Northern California Institute for Research and
Education. In the past, ADNI has also received funding from the National
Institute of Biomedical Imaging and Bioengineering, the Canadian
Institutes of Health Research, and private sector contributions through
the Foundation for the National Institutes of Health (FNIH) including
generous contributions from the following: AbbVie, Alzheimer's
Association; Alzheimer's Drug Discovery Foundation; Araclon Biotech;
BioClinica, Inc.; Biogen; BristolMyers Squibb Company; CereSpir, Inc.;
Cogstate; Eisai Inc.; Elan Pharmaceuticals, Inc.; Eli Lilly and Company;
EuroImmun; F. Hoffmann-La Roche Ltd and its affiliated company
Genentech, Inc.; Fujirebio; GE Healthcare; IXICO Ltd.; Janssen Alzheimer
Immunotherapy Research \& Development, LLC.; Johnson \& Johnson
Pharmaceutical Research \& Development LLC.; Lumosity; Lundbeck; Merck
\& Co., Inc.; Meso Scale Diagnostics, LLC.; NeuroRx Research; Neurotrack
Technologies; Novartis Pharmaceuticals Corporation; Pfizer Inc.; Piramal
Imaging; Servier; Takeda Pharmaceutical Company; and Transition
Therapeutics.

The A4 and LEARN Studies were supported by a
public-private-philanthropic partnership which included funding from the
NIA of the NIH (R01 AG063689, U19AG010483 and U24AG057437), Eli Lilly
(also the supplier of active medication and placebo), the Alzheimer's
Association, the Accelerating Medicines Partnership through the
Foundation for the National Institutes of Health, the GHR Foundation,
the Davis Alzheimer Prevention Program, the Yugilbar Foundation, Gates
Ventures, an anonymous foundation, and additional private donors to
Brigham and Women's Hospital, with in-kind support from Avid
Radiopharmaceuticals, Cogstate, Albert Einstein College of Medicine and
the Foundation for Neurologic Diseases. Open access funding provided by
SCELC, Statewide California Electronic Library Consortium.

\bibliography{references.bib}

\end{document}